Title:      Theoretical aspects of Vertical and Lateral Manipulation of Atoms
Authors:    C. Ghosh, A. Kara, T.S. Rahman

Address:    Department of Physics
            116 Cardwell Hall
            Kansas State University
            Manhattan, KS 66506
            USA
Phone:      +1-785-532-6786
Fax:        +1-785-532-6806
Email:      **rahman@phys.ksu.edu**

# Theoretical aspects of Vertical and Lateral Manipulation of Atoms

Chandana Ghosh[1,2], Abdelkader Kara[2] and Talat S. Rahman* [1,2]

[1]*Fritz-Haber-Institut der Max-Planck-Gesellschaft, Faradayweg 4-6, D-14195 Berlin, Germany*
[2]*Department of Physics, Cardwell Hall, Kansas State University, Manhattan, KS 66506, USA*
*(\* Corresponding author: Fax: +1-785-532-6806; email: rahman@phys.ksu.edu)*

## Abstract

Using total energy calculations, based on interaction potentials from the embedded atom method, we show that the presence of the tip not only lowers the barrier for lateral diffusion of the adatom towards it, but also shifts the corresponding saddle point. For a Cu adatom at a (100) microfacetted step on Cu(111) this shift is $0.6 \overset{o}{A}$. The effect of the tip geometry and shape on the energetics of lateral manipulation was found to be subtle. In the case of vertical manipulation of a Cu adatom on flat, stepped, and kinked Cu surfaces we find an unusual but interesting result. It is found that as the tip approaches the surface, it becomes easier to extract the adatom from the stepped and the kinked surfaces, as compared to the flat surface. This counter intuitive result can be explained in terms of tip induced changes in the bonding of the adatom to its low coordinated surroundings.

# I. Introduction

One of the spectacular scientific advances of recent times is the ability to move atoms and molecules, in a controlled manner, using the tip of a Scanning Tunneling Microscope (STM) [1-4]. Individual atoms and molecules have been manipulated to move both laterally and vertically by this instrument in a number of research laboratories [1-7]. This technique opens the way for a variety of technological applications including nanostructuring of surfaces and manipulation of chemical reactions – both of which are important to the fruition of some of the expectations from nanotechnology. While for several types of vertical manipulation the picking up of an atom at one point and dropping it off at another, on the atomic scale, depends on the bias voltage that is applied, for lateral manipulation the magnitude of the electric field is weak [7]. These experiments have raised several questions. Can lateral manipulation be performed by an Atomic Force Microscope? In what precise way does the presence of the tip alter the potential energy surface? How does the vicinity of a step-edge or kink-site affect the vertical manipulation of an adatom? Do the atomic – scale characteristics of the tip make a difference? The answers to a few of these questions have already been provided in previous theoretical studies [8,9]. It appears from these studies that most of the characteristics of manipulation on metal surfaces can be explained through considerations of the system energetics dictated by interatomic forces alone. These calculations naturally take into account some of the changes induced by the tip on the potential energy surface available to the adatom. As we shall see, more detailed examination of the potential energy surface, however, reveals effects of the tip that were not uncovered in these earlier studies. The goal of the present paper is to provide answers to these and related questions, which provide greater insights into the changes in the potential energy surface induced by the presence of the tip. While not the subject of direct investigation, these detailed examinations of the potential energy surfaces set the stage for the calculations of the vibrational frequencies of adsorbates on surfaces in the presence of the STM tip. As in Ref. [8], our model system has been inspired by the experimental work carried out earlier by Meyer et al. [14,15].

Since it is not yet computationally feasible to carry out calculations of the type we have in mind with first principles electronic structure methods [16], we have employed interaction potentials from the embedded atom method (EAM) [10] to examine the energetics of atom manipulation on metal surfaces. Although these interaction potentials are semi-empirical, we have had remarkable success in calculating the structural and dynamical properties of both flat and stepped surfaces of Cu, Ag and Ni using these potentials [17,18]. For lateral manipulations we have mapped out the potential energy surface in the presence of the tip using a fine grid. It is the usage of this fine grid which sets this calculation apart from that in Ref. 8 and reveals to us the shifts in the saddle points in the diffusion paths of adatoms in the presence of the tip. This result, which was not anticipated in earlier work [8], provides a more realistic estimate of the activation energy barriers for adatom diffusion and manipulation. Additionally, the present work takes a comparative look at the energetics involved in the vertical manipulation of atoms on flat, stepped and kinked Cu surfaces. In Section II, we present a brief description of the model system. This is followed by a brief account of the theoretical techniques employed, in Section III. Section IV contains results for lateral manipulation, while those for vertical manipulation are summarized in Section V. Conclusions are presented in Section VI.

## II. Model Systems

The model system for the manipulation process consists of two parts: the adatom on a metal substrate, and the tip, which is a nanostructure itself. We take the substrate to be a slab consisting of 8 atomic layers in fcc stacking with (111) orientation and 10x12 atoms in each layer (12 chains of atoms per layer). We have performed calculations on larger systems and arrived at the conclusion that the above dimensions are sufficient to avoid problems arising from the finite size of the simulation cell. The (100) microfacetted step and the kinked surfaces are created by deleting 7 and 7.5 atomic chains respectively from the surface layer. Lateral manipulation is carried out on the stepped surface only, while vertical manipulation is performed on all three types of surfaces, for comparison. Both sharp and blunt tips having either fcc(100), or fcc(111) crystal geometry are used in the simulations. Sharp tips consist of 35 metal atoms with 1 atom at the tip apex (Fig. 1a). The blunt tips have 34 atoms, leaving the (100) and (111) tips with 4 and 3 atoms at the apex, respectively. The total height of the blunt and sharp tips are thus 8.32 Å and 10.40A, respectively. Since the interactions potentials for the materials considered are short ranged, the sizes of the tips chosen here are sufficient to examine their effect on the potential energy surface in the vicinity of the adatom.

## III. Theoretical Method

As in a previous study [8], we restrict ourselves to an examination of the energetics of the system using reliable many body interaction potentials from the embedded atom method (EAM) [10], with parameterization by Voter and Chen [11]. The total energy of the system consisting of the tip, the adatom and the substrate (Fig. 1a) is minimized using the conjugate gradient method [12]. Dynamical effects induced by atomic vibrations and contribution of the vibrational entropy to activation free energies would play a role with increasing surface temperature and if the time scales of the motion induced by the tip were comparable to those of the vibrations of the system. Since the experiments that have motivated our calculations were performed at temperatures around 50K and the time scales for the atomic motion is about 1ps, we do not expect surface vibrations to play a significant role in the manipulation processes that we examine here. Also the zero point motion of atoms like Cu is not significant to change the qualitative behavior of these systems. Van der Waals forces are also not expected to be predominant [13], since the tip radius is taken to be just a few angstroms and the vertical separations between the tip and the adatom range from 3.0Å to about 7.5Å.

During the energy minimization calculations, the four atoms at the corners of the stepped surface, as well as, those at the two corners of the lower terrace, and all atoms of the last two layers of the slab are kept rigid. All tip atoms, except for the ones in the top two layers, are also allowed to relax to their equilibrium positions for each simulation. The results quoted in this paper are for the fully relaxed tip and surface systems. In the case of vertical manipulation the adatom is constrained in the Z direction (normal to the surface). This ensures that the adatom does not fall back to the surface when we raise it in small intervals from the surface to the tip in the energy minimization process. The plots of the total energy versus the Z coordinate of the adatom are obtained for several tip heights.

For lateral manipulation the total energy is calculated for all configurations of the system with the adatom placed on each point along the fine 10Å x 10Å grid consisting of 100 x 100 points and covering parts of the regions of the two terraces adjoining the step. From the total energy of the system for each of the $10^4$ configurations, the generated 3D potential energy surface of the system, for a tip placed directly above the adatom and at a height of 4.75Å from the surface (about 2.7 Å above the adatom) is shown in Fig. 1b. The data has been smoothened to a small degree using the Gaussian smoothing technique [12]. The contour plot in Fig. 1c, which is a 2D top view of the same potential energy surface, has been obtained through the application of a graphics package (Xfarbe) [20]. In this plot the perturbation caused by the presence of the tip is seen on the top central portion (the tip is directly above that region). In regions away from the tip, the usual symmetry of the surface is reflected. The x and y axes have a scales from 0 to 100 indicating the 100x100 points on the 10Å x10Å grid. From the calculated potential energy surface, the activation energies for the adatom to diffuse in various directions are easily obtained.

## IV. Results

In this section we discuss the results of the different aspects of manipulation, lateral and vertical. We examine the effects of the tip geometry, shape, and height from the substrate on the characteristics of the manipulation process. As expected beyond a certain separation between the tip and the substrate, the tip itself gets distorted. We report here results only for those tip heights for which the tip atoms remain intact.

To establish the importance of the usage of the grid, we consider the lateral manipulation of a Cu adatom at a (100)-microfacetted Cu(111) step face using a sharp Cu(111) tip, for direct comparison with results from Ref. [8]. The plot in Fig. 2a shows that the qualitative features of the calculated energy barriers for an adatom to diffuse in the direction of the tip, and away from it (opbarrier), are similar to those in Ref. [8]. However, in the present work the minimum barrier is not zero. Instead it is 146.6 meV and is reached at a tip-adatom lateral separation of about 2.0Å (Table 1) and a vertical separation of 2.75Å. We observe that sharp Cu(111) tips approaching the adatom vertically any closer than 2.75Å get distorted due to the attractive forces between the surface and the tip. Such distortion of the tip was also found in Ref. [9]. The 2D and 3D contour plots in Figs. 1b and 1c clearly show that when the tip is far away from the substrate the system is not perturbed and the potential energy surface for the adatom has the substrate symmetry. The region in the vicinity of the tip shows a perturbation and the saddle point shifts with respect to its original unperturbed position. This is why the diffusion barriers calculated with respect to the actual saddle point are different from those in Ref. [8] in which the saddle point was assumed to be at the bridge site (see Fig. 3). The shift in the saddle point is clearly marked in the Fig. 1c.

To examine the effect of the tip geometry on lateral manipulation, we present in Fig. 2b the results for a Cu tip of (100) geometry. These results show the Cu(100) tip to be more effective in lateral manipulation as compared to the Cu(111) tip, as the barrier here is lowered to 61.7 meV. Figures 2 also include the cases of manipulation with blunt Cu(100) and Cu(111) tips. For the case of a blunt (111) tip, the lowest tip-adatom vertical separation achieved without distorting the tip is 3.0Å. For this case, the lowest barrier for the adatom to diffuse towards the tip is found to be 143 meV at a lateral separation of about 2.0Å. This value should be compared with the negative barrier obtained in Ref. [8], whose implication would have been a preference of the adatom for the bridge site rather than the next hollow site. This would have meant that a lateral motion would not take place at all. The usage of the denser grid is thus found to provide better sampling of the potential energy surface. Comparison of the barriers in Fig. 2a and 2b shows the blunt Cu(100) tip to be effective in reducing the barrier over a larger lateral separation range than the sharp tip. In Table 1 we have summarized our calculated activation barriers for lateral manipulation with sharp and blunt tips with either (111) or (100) geometry.

## V. Vertical Manipulation

Vertical manipulation is generally more difficult as compared to lateral manipulation because the activation energy barriers needed to lift the adatom from the surface may be substantially higher than those for lateral manipulation. However, a careful study shows that during the process of bringing the tip closer to the adatom, the barrier to jump to the tip does vanish at a certain height (tip-substrate distance) and the adatom can be pulled to the tip apex [5,6]. We present in Figs. 4a and 4b the results obtained for a blunt Cu(100) tip, as we have found it to be more effective than a sharp tip in vertical manipulation. Figure 4a shows that as the tip is lowered the energy profile of the system changes from a double-well like curve, to one in which the barrier hump disappears and the two wells merge into a single one. In other words, the left minimum of the curves (which is the position of adatom near the surface) merges with the right minimum (adatom at tip apex), showing that the energy barrier for the adatom to be pulled up by the tip goes to zero at a certain tip height.

Interestingly, for the kinked and the stepped surfaces we find a "floating region" (at a tip height between 7.5Å - 6.5Å), in which the total energy of the system is a minimum at a point between the surface and the tip apex. This is reflected in Fig. 4a for adatom manipulation on the stepped Cu(111) surface. In this region the adatom is equally attracted to the surface and the tip and may float in between. For the adatam on the flat surface such a region is not conspicuous.

A plot of the energy barriers versus the inverse of the tip height, in Fig. 4b, shows that at large tip heights, the barrier to pull an adatom from a flat surface is lower than that from a stepped or a kinked surface as in expected from considerations of the local coordination of the adatom on these surfaces: the adatom needs to break 3 bonds in the case of the flat, 5 for the stepped, and 6 for the kinked surface, to detach from them. However, as the tip is lowered its impact on the potential energy surface is more pronounced on the stepped and kinked surfaces than the flat one and it becomes easier to extract an atom from a stepped or a kinked surface. This crossover point at which the decrease in barriers becomes more rapid for the stepped and the kinked surfaces as compared to the flat surface, is clearly seen in the Fig. 4b. This unexpected result implies that the presence of the tip not only weakens the bonding of the adatom, but it also highly affects the local environment of the adatom, especially in the case of the low-coordinated surroundings of the stepped and the kinked surfaces.

## VI. Conclusions

In summary, we have shown that in the presence of a tip, the saddle point for an adatom to diffuse laterally from one hollow site to the next, along a (100)-microfaceted step edge on Cu(111), shifts away from the bridge site. The activation barrier for the adatom to hop towards the tip is considerably lowered but does not vanish. The lowest barrier is reached at a lateral separation between 2.0Å -3.0Å. Our results show that the qualitative changes in the potential energy surface are independent of the shape and geometry of the tip but we see a quantitative difference when considering the tip details. Although a sharp tip lowers the barrier slightly more than a blunt tip, the effect of the blunt tip extends over a larger range of lateral separations, which may be traced to the higher coordination it offers to the adatom.

We have also found that in the presence of the tip, the neighboring atoms relax towards the tip whereas the tip apex atom relaxes upwards into itself [19]. At smaller vertical separations between the tip and the adatom (2.5Å), tip apex is found to relax downward toward the surface, showing a mutual attractive force. From the results for vertical manipulation we find that a blunt tip is most effective in lowering the barrier to zero as it gradually comes closer to the adatom. At large (>12Å) tip heights the energy required to pull an adatom from a flat surface (2.47eV) is lower than that from a stepped (3.21eV) and a kinked surface (3.53eV). As the tip is lowered to facilitate manipulation, the changes in the surface energetics are such that it is easier to extract an adatom from a stepped or a kinked surface than from a flat one, implying that the tip affects the low coordinated local environment of the adatom, which assists in its extraction. In addition, for specific tip heights, there is a floating region on the kinked and stepped surfaces in which the adatom is found to be equally attracted to the substrate and to the tip apex atoms.

**Acknowledgements**

This work was partially supported by DOE-EPSCoR/K-TECH under Grant No. DE-FG02-98ER45713. Computations were performed on the HP Exemplar system funded partially under NSF Grant No. CDA-9724289. One of us (TSR) thanks the Alexander von Humboldt Foundation for the award of the Forschungspreis, which also facilitated this work.

**Table 1** Comparison of energy barriers for lateral manipulation on a stepped Cu(111) surface with a (100) step microfacet, for various types of tips.

| Tip geometry | Shape | Tip-adatom separation (Å) | Activation Barrier (meV) | Lateral separation(Å) |
| --- | --- | --- | --- | --- |
| Cu(100) | Sharp | 2.75 | 61.7 | 2.55 |
| Cu(111) | Sharp | 2.75 | 146.9 | 2.00 |
| Cu(100) | Sharp | 3.50 | 203.9 | 2.00 |
| Cu(111) | Sharp | 3.50 | 223.2 | 2.55 |
| Cu(100) | Blunt | 3.00 | 100.5 | 2.55 |
| Cu(111) | Blunt | 3.00 | 141.3 | 2.00 |
| Cu(100) | Blunt | 3.50 | 199.1 | 3.05 |
| Cu(111) | Blunt | 3.50 | 202.9 | 2.00 |

# Figure Captions

**Fig. 1a** Model System for lateral manipulation. Shown here is the grid along which the adatom is placed.

**Fig. 1b** 3D plot of the PES when the tip is 2.75Å above the adatom which is along the step edge.

**Fig. 1c** Top view of the contour plot of the total energy of the system, showing the perturbation due to the tip (upper middle part). The long dark line indicates the direction along the step edge (reaction coordinate). The small line shows the shift in the position of the saddle point (⊗) with respect to its ideal position (which is where the dashed line meets the small line). The circles indicate the minima (fcc and hcp sites), the crosses, the maxima (as can be seen, these are the atop sites) and the circles with crosses, the saddle points (bridge sites).

**Fig. 2a** Barrier (filled circles) and op-barrier (open circles) for (100) and (111) sharp tips for a tip apex-adatom separation of 2.75Å. Solid lines are for the (100) tip and the dashed lines are for the (111) tip.

**Fig. 2b** Barrier (filled circles) and op-barrier (open circles) for (100) and (111) blunt tips for a tip-adatom separation of 3.0Å. Solid lines are for the (100) tip and the dashed lines are for the (111) tip.

**Fig. 3** Broken symmetry due to the tip. Subtracting the energies at points **Hollow1** and **B** would yield a zero barrier, but considering the shifted barrier **S** and subtracting the energies at points **Hollow1** and **S** gives a positive barrier $E_b$.

**Fig. 4a** Vertical Manipulation – disappearance of the barrier for a stepped Cu(111) surface as the tip is lowered towards the adatom. The floating region is also seen.

**Fig. 4b** Comparison of the barriers for the flat, stepped and kinked Cu(111) surfaces with respect to the inverse of the tip height.

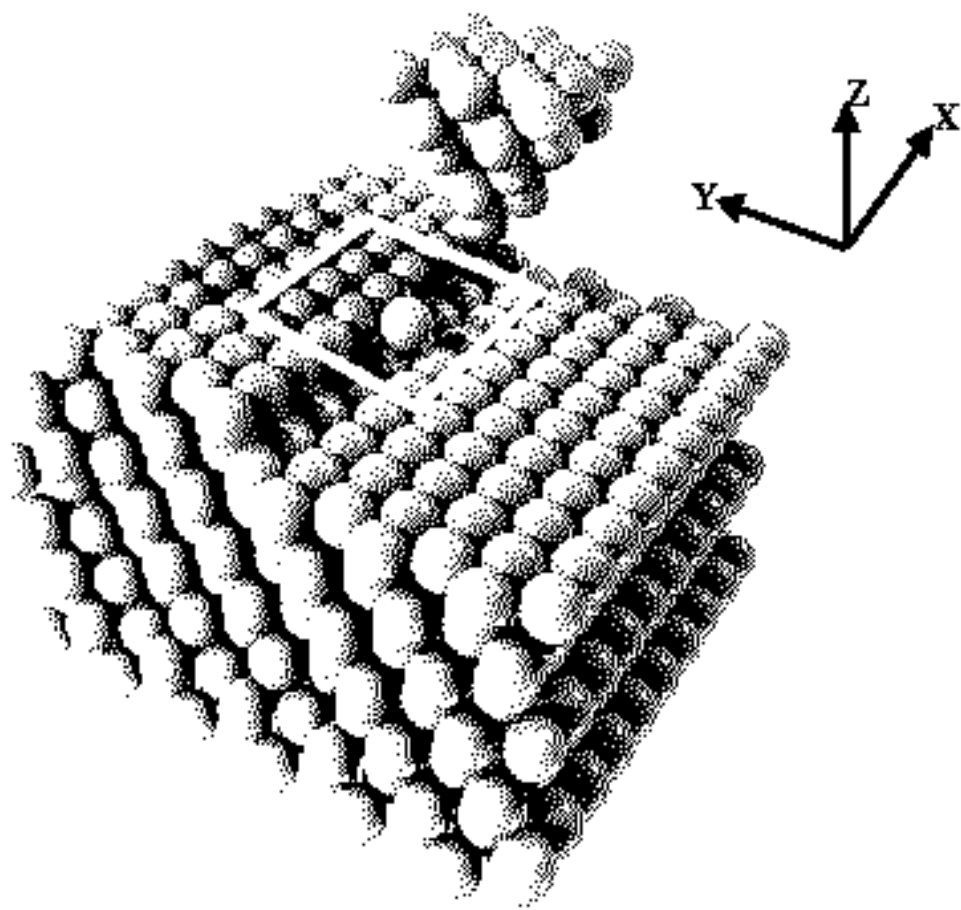

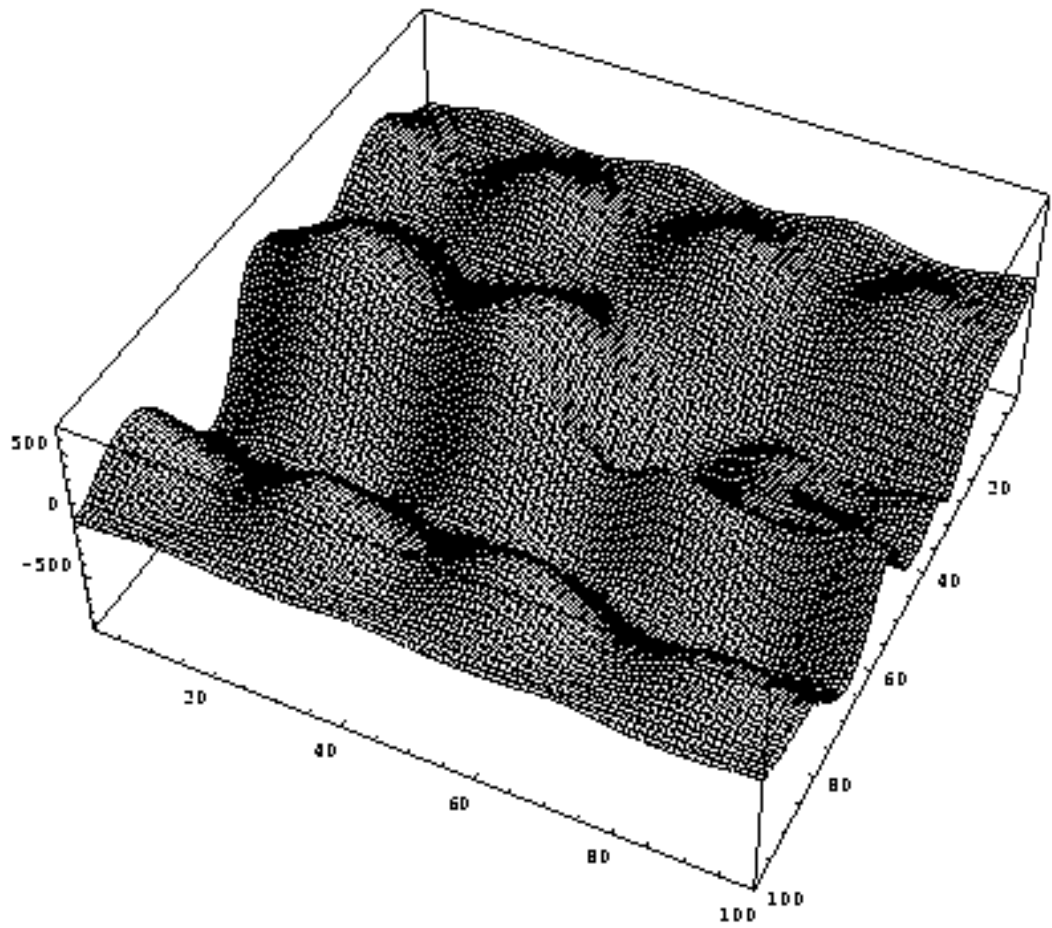

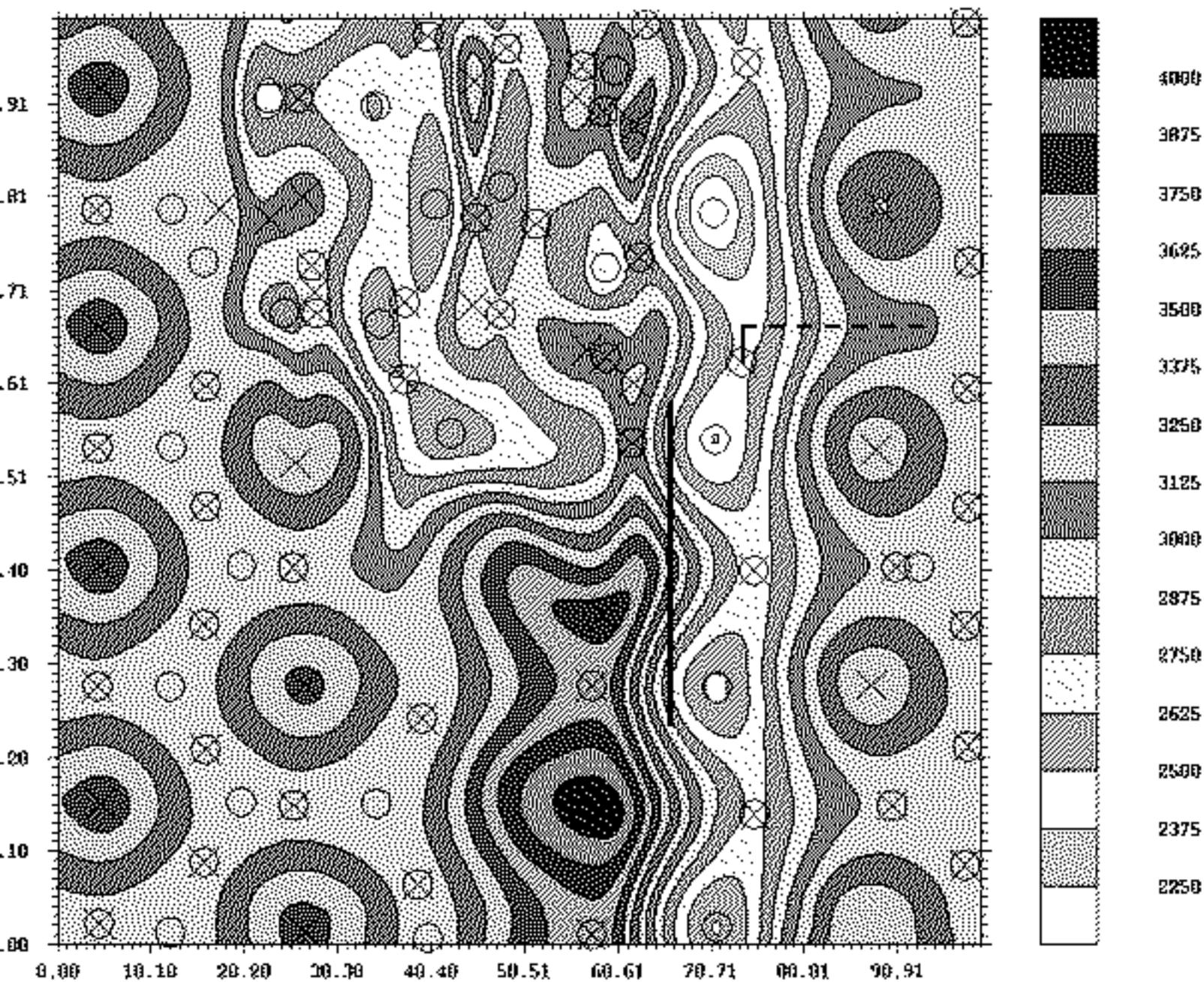

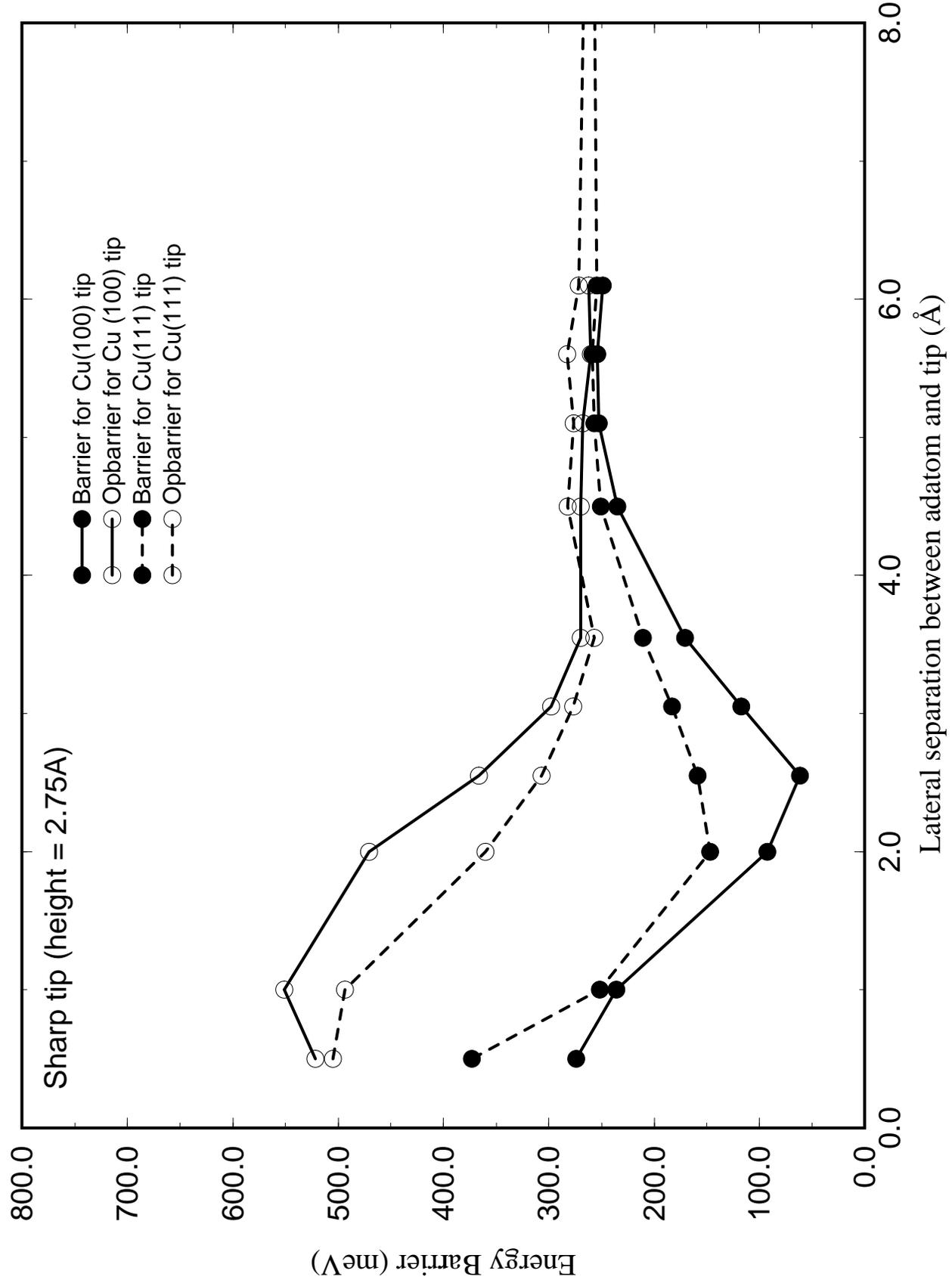

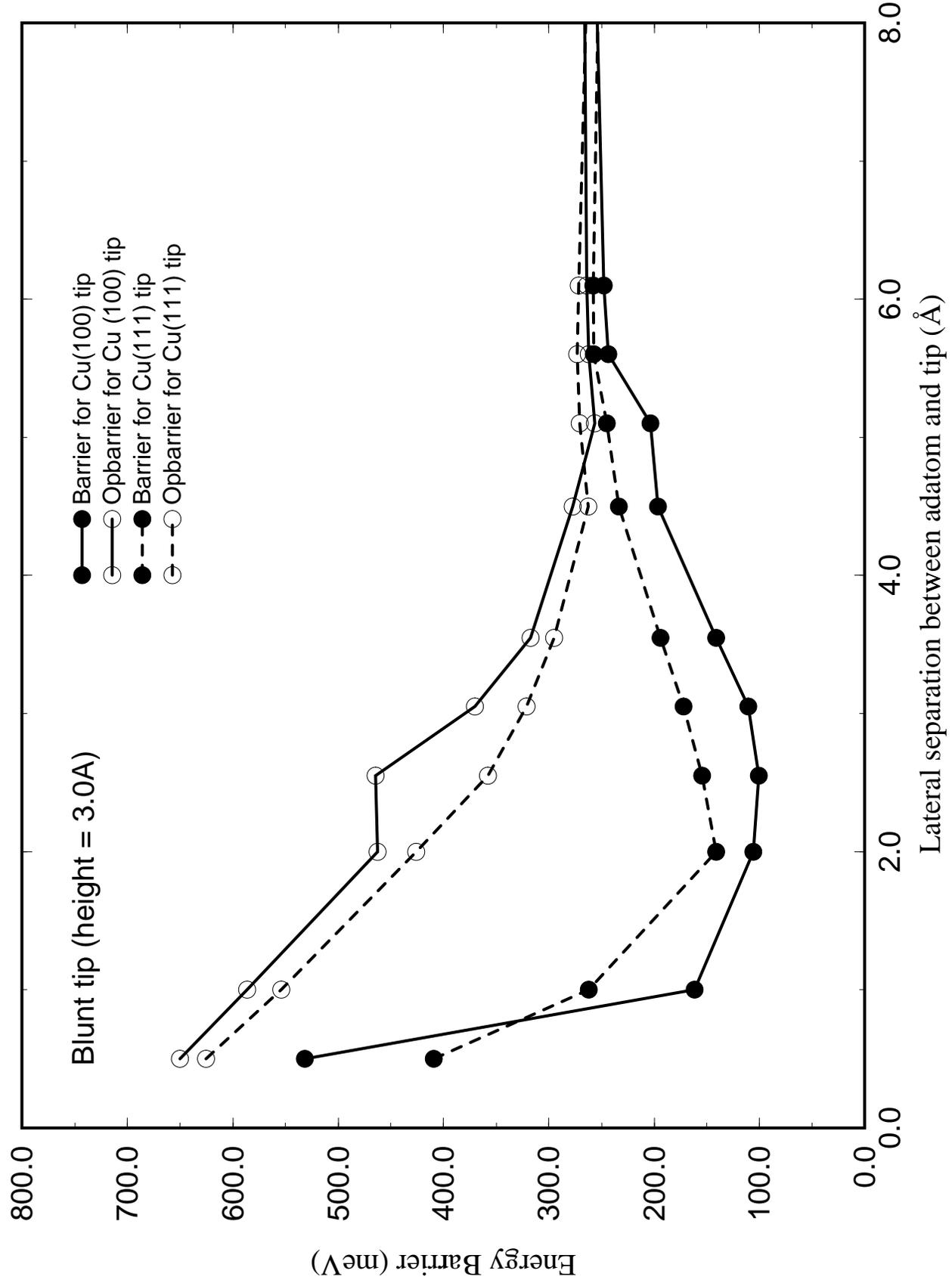

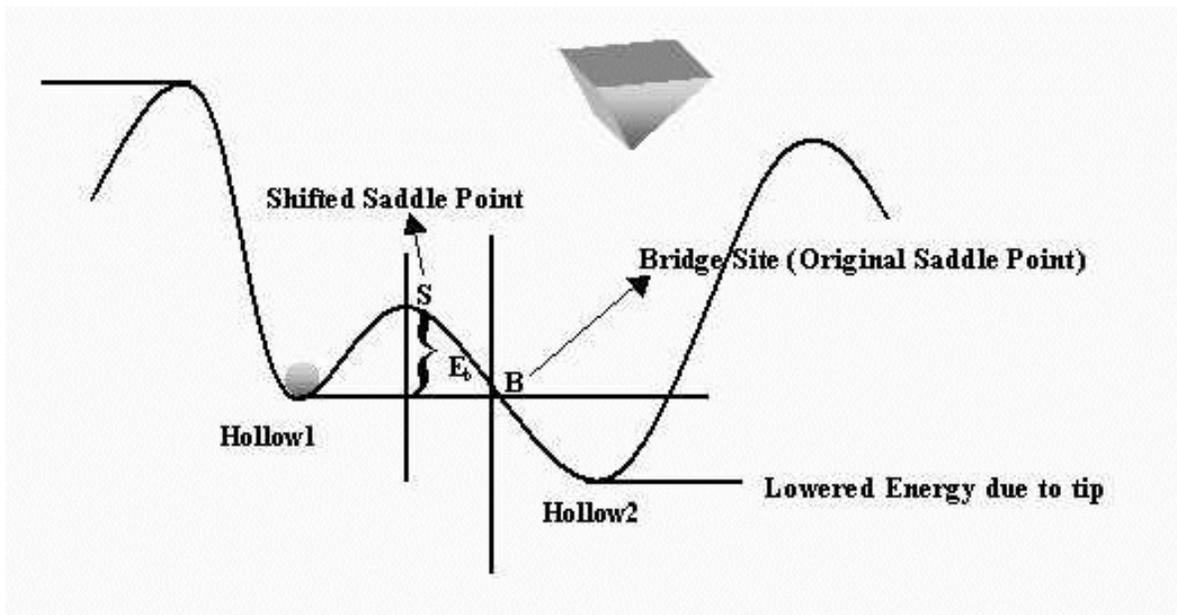

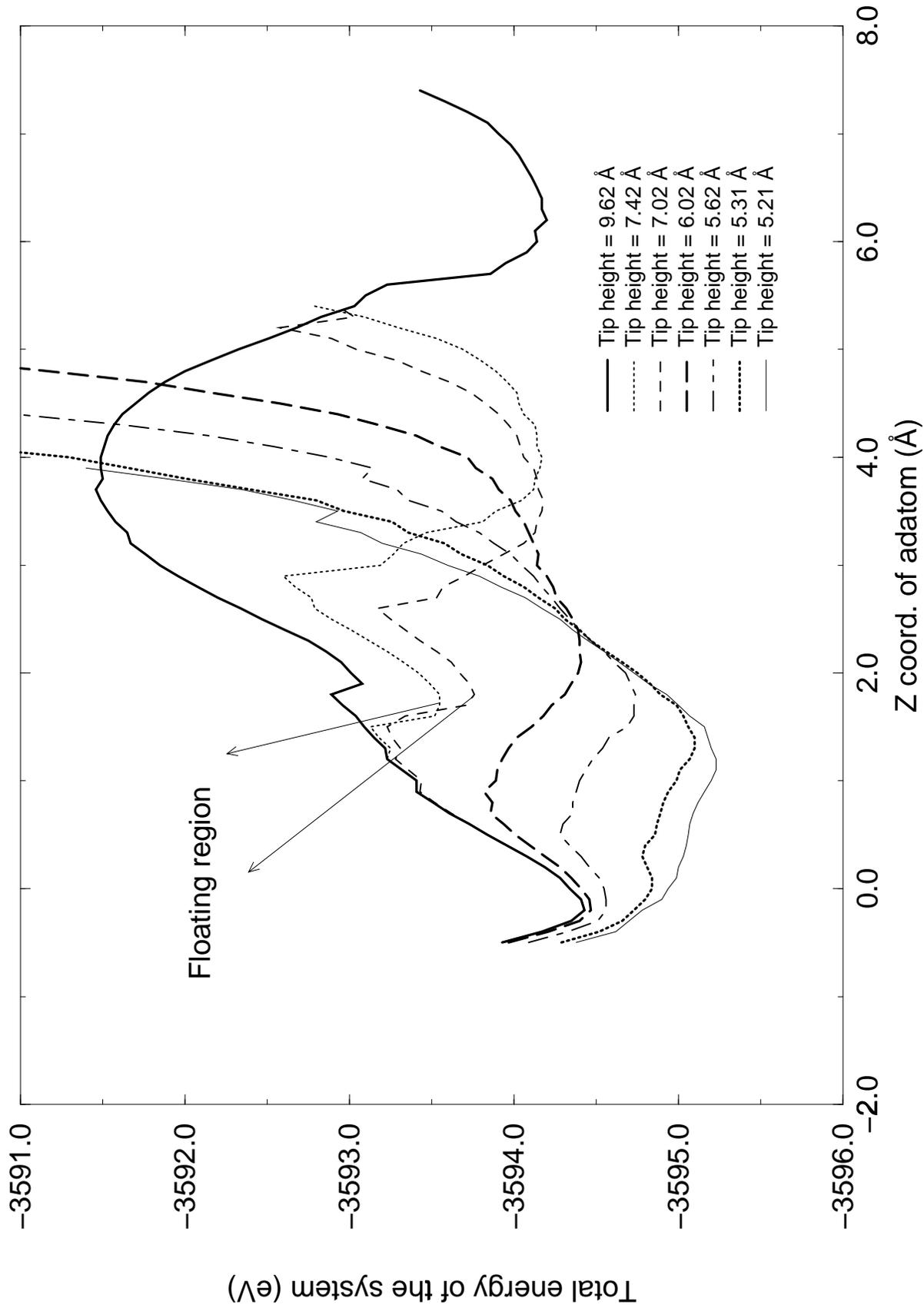

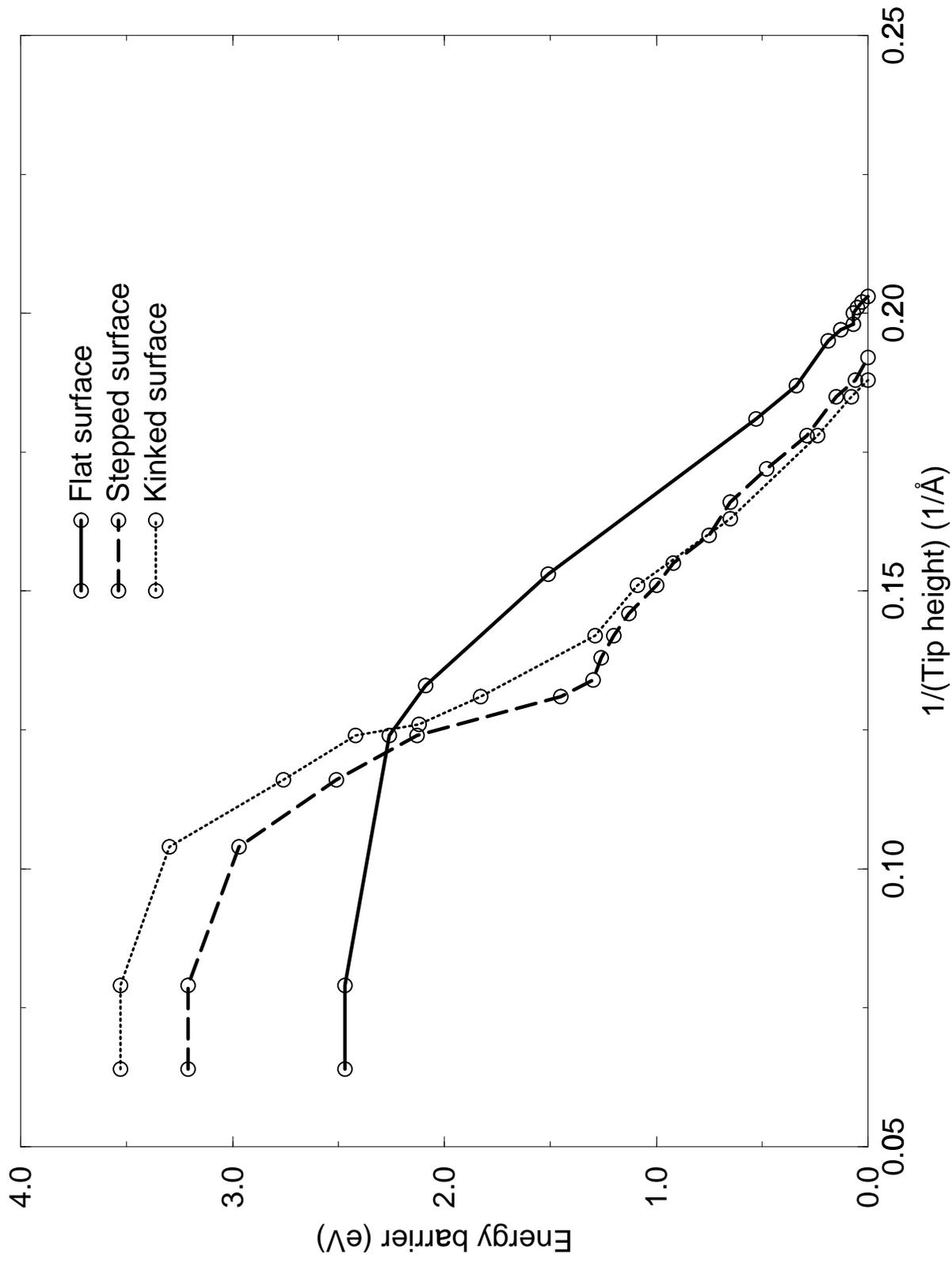